\documentclass[12pt]{iopart}
\usepackage{graphicx}

\begin{document}

\textheight=8.25in

\title[The scalar curvature in Regge calculus]{A geometric
  construction of the Riemann scalar curvature in Regge calculus}

\author{Jonathan R. McDonald and Warner A. Miller}

\address{Department of Physics, Florida Atlantic
         University, Boca Raton, FL 33431, USA} 
 
\ead{wam@fau.edu}
        
\begin{abstract} 
  The Riemann scalar curvature plays a central role in Einstein's
  geometric theory of gravity.  We describe a new geometric
  construction of this scalar curvature invariant at an event (vertex)
  in a discrete spacetime geometry.  This allows one to constructively
  measure the scalar curvature using only clocks and photons. Given
  recent interest in discrete pre-geometric models of quantum gravity,
  we believe is it ever so important to reconstruct the curvature
  scalar with respect to a finite number of communicating observers.
  This derivation makes use of a new fundamental lattice cell built
  from elements inherited from both the original simplicial (Delaunay)
  spacetime and its circumcentric dual (Voronoi) lattice. The
  orthogonality properties between these two lattices yield an
  expression for the vertex-based scalar curvature which is
  strikingly similar to the corresponding hinge-based expression in
  Regge calculus (deficit angle per unit Voronoi dual area).  In
  particular, we show that the scalar curvature is simply a
  vertex-based weighted average of deficits per weighted average
  of dual areas.
\end{abstract}


\submitto{\CQG}

\maketitle

The Riemann scalar curvature invariant plays such a central role in
Einstein's still standard geometric theory of gravitation introduced
in 1915, its centrality in the theory cannot be over emphasized.  The
extremum of this quantity over a proper 4-volume of spacetime, yields
a solution compatible with Einstein's field equations. It is this
scalar, so central to the Hilbert action, which yields the
conservation of energy-momentum (contracted Bianchi identities) when
variations are done with respect to the diffeomorphic degrees of
freedom of the spacetime geometry.  This scalar also augments the
Ricci tensor in coupling the non-gravitational fields and matter to
the curvature of spacetime. It not only appears in its 4-dimensional
form in the integrand Hilbert action principle of general relativity,
it makes its presence felt in 3-dimensions as an ``effective potential
energy'' in the ADM action.

Given this curvature invariant's pivotal role in the theory of general
relativity, we believe it is important to understand how to locally
construct this geometric object at a chosen event in an arbitrary
curved spacetime. Given recent interest in discrete pre-geometric
models of quantum gravity, it is ever so important to reconstruct the
curvature scalar with respect to a finite number of observers and
photons\cite{L06,AJL06}. Even though we do have familiar discrete
representations of each of the twenty components of the Riemann
curvature tensor in terms of geodesic deviation or parallel transport
around closed loops\cite{S60,B66,CD86}, and apart from the sterile act
of simply taking the trace of the Riemann tensor, we are not aware of
such a chronometric construction of the scalar curvature.

In this manuscript we provide such a discrete geometric description of
this scalar curvature invariant utilizing the approach of Regge
calculus\cite{R61,MTW73,TW92}, and the convergence-in-mean of Regge
calculus was rigorously demonstrated\cite{CMS84}. In the spirit of
quantum mechanics and recent approaches to quantum gravity, our
construction uses only clocks and photons local to an event on an
observer's world line. Furthermore, this construction is based on a
finite number of observers (clocks) exchanging a finite amount of
information via photon ranging and yields the scalar curvature
naturally expressed in terms of Voronoi and Delaunay
lattices\cite{book}.  It has been shown that these lattices naturally
arrise in Regge
calculus\cite{B86,CDM89,FL84,FFLR84,CFL82,CFL82b,L83,HW86,HW86b,M86,M97}.This
constriction further emphasizes the fundamental role that Voronoi and
Delaunay lattices have in the discrete representations of spacetime
which is perhaps not so surprising given its preponderant role in
self-evolving and interacting structures in nature\cite{book}. In this
analysis we introduce a new hybrid (half Voronoi, half Delaunay)
simplex which we argue is fundamental to Regge calculus\cite{M97} and
perhaps fundamental to any discrete representation of classical and
quantum gravity.

Consider the familiar simplical representation of the geometry of
spacetime common in Regge calculus\cite{R61,MTW73}.  Here the
spacetime is composed of a countable number simplicies.  Each
4-simplex is endowed with a flat Minkowski spacetime interior. This
lattice is a 4-dimensional spacetime Delaunay lattice.  By
construction, the curvature in this lattice spacetime does not reside
in its 4-simplicies, nor in its tetrahedra; however, the curvature is
concentrated on each of its 2-dimensional triangle hinges, $h$.  Each
of these hinges is the meeting place of three or more 4-simplicies. In
the traditional description of Regge calculus, this hinge-based
curvature is viewed as a conic singularity; however, it has been shown
that the areas $h^*$ of the Voronoi lattice dual to the Delaunay
simplicial lattice provides a natural area to distribute the
curvature\cite{M97,M86}. The Voronoi lattice is constructed in the
usual way by utilizing the circumcentric dual of the Delaunay
lattice\cite{book}.

The key to our derivation of the Riemann-scalar curvature is the
identification $I_h \equiv I_v$ of the usual hinge-based expression
the Regge calculus version of the Hilbert action principle
\cite{R61,M97} with its corresponding vertex-based expression. We
begin with the Hilbert action in a $d$-dimensional continuum
spacetime, which is expressible as an integral of the Riemann scalar
curvature over the proper $d$-volume of the spacetime.
\begin{equation}
I = \frac{1}{16 \pi} \int R\, dV_{proper}
\end{equation}
On our lattice spacetime, and following the standard techniques of
Regge calculus, we can approximate this action as a sum over the
triangular hinges $h$.
\begin{equation}
\label{Ih}
I \approx I_h =   \frac{1}{16 \pi} \sum_{hinges,\ h} R_h \Delta V_h
\end{equation}
Here, $R_h$ is the scalar curvature invariant associated to the hinge,
and $\Delta V_h$ is the proper 4-volume in the lattice spacetime
associated to the hinge $h$. Following earlier work by the
authors\cite{M97}, this curvature will be defined explicitely below.
Though, non-standard in Regge calculus, we may also express the action
in terms of a sum over the vertices of the simplicial $d$-dimensional
Delaunay lattice spacetime.
\begin{equation}
\label{Iv}
I \approx  I_v = \frac{1}{16 \pi} \sum_{vertices,\ v} R_v \Delta V_v
\end{equation}
It is the Riemann scalar curvature ($R_v$) at the vertex $v$ that
appears in this expression that we seek in this manuscript, and it is
the equivalence between (\ref{Ih}) and (\ref{Iv}) that will yield
it. But first we must use the orthogonality inherent between the
Voronoi and Delaunay lattices to determine the relevant 4-volumes
($\Delta V_v$ and $\Delta V_h$).\footnote{Although the primary concern
  of the authors is to apply these results to the 4-dimensional
  pseudo-Riemannian geometry of spacetime, our equations are valid for
  any Riemann geometry of dimension $d$. Therefore, in the text and
  equations to follow we will explicity use the symbol $d$ to
  represent the dimensionality of the geometry, the reader interested
  in general relativity can simply set $d=4$. }

\begin{figure}
\centering
\includegraphics[width=0.8\linewidth]{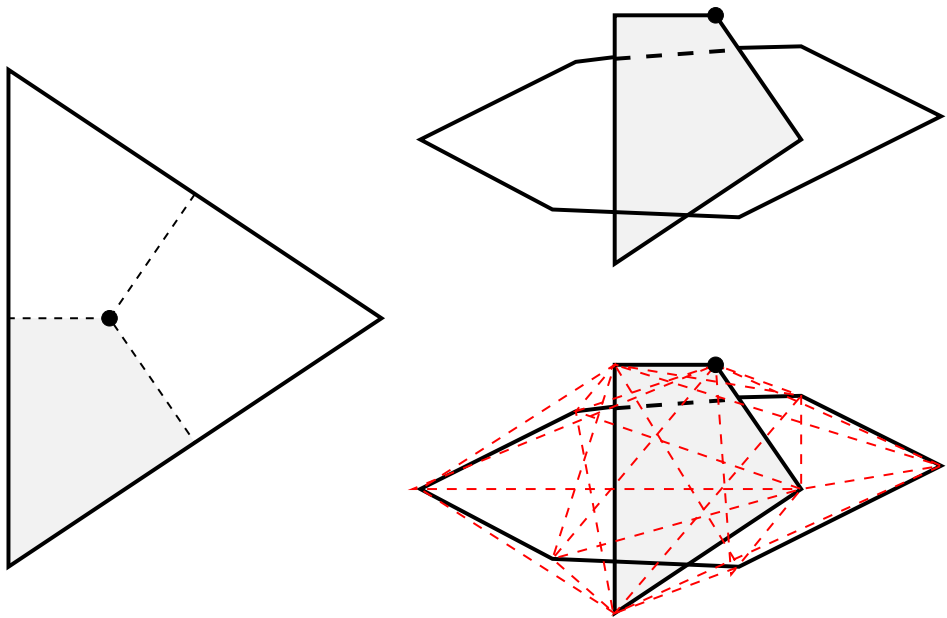}
\caption{The triangle hinge $h$ to the left is partitioned into three
  areas. The shaded region ($A_{hv}$) represents the portion of the
  triangle that is closer to the lower vertex than its other two
  vertices. The darkened and pronounced vertex appearing in each of
  the three line drawings of this figure is the circumcenter of the
  hinge, $h$. Each hinge has its corresponding 2-dimensional dual
  Voronoi area ($A^*_h$) shown in the upper right part of the figure
  as a pentagonal shaped polygon, and illustrate this dual area as
  ``encircling'' the $d$-dimensional ``kite'' hinge.  In bottom right
  portion of the figure, the ``kite'' hinge is connected to its dual
  Voronoi polygon by ($4\times 6 = 24$) new lattice edges -- thus
  forming the Voronoi-Delaunay hybrid d-cell which is fundamental to
  our derivation, the derivation of the Hilbert action in Regge
  calculus, and, we believe, fundamental to any discrete
  representation of gravitation. Each if these edges, as well as the
  edges of the Voronoi area are algebraic functions of the original
  Regge simplicial (Delaunay) lattice spacetime, and accordingly we
  have not added or subtracted any degrees of freedom. These new
  hybrid d-cells provide a new, and proper tiling of the lattice
  spacetime. }
\label{fig:ahv}
\end{figure}

Consider a vertex $v$ in the Delaunay lattice, and consider a triangle
hinge $h$ having vertex $v$ as one of its three corners. We define
$A_{hv}$ to be the fraction of the area of hinge $h$ closest to vertex
$v$ than to its other two vertices ( Figure~{\ref{fig:ahv}}). Dual to each
triangle hinge, and in particular to triangle $h$, is a unique
co-dimension 2 area, $A^*_h$, in the Voronoi lattice. This area
necessarily lies in a $(d-2)$-dimensional hyperplane orthogonal to
the 2-dimensional plane defined by the triangle $h$. The number of
vertices of the dual $(d-2)$-polygon, $h^*$, is equal to the number of
$d$-dimensional simplicies hinging on triangle $h$, and is always
greater than or equal to three.  If we join each of three vertices of
hinge $h$, with the all of vertices of $h^*$ with new edges, then we
naturally form a $d$-dimensional proper volume associated with a
vertex $v$ and hinge $h$. This $d$-dimensional polytope is a
hybridization of the Voronoi and Delaunay lattices, they completely
tile the lattice spacetime without gaps or overlaps, and they inherent
their rigidity from the underlying simplicial lattice.
\begin{equation}
\label{Vhv}
\Delta V_{hv} \equiv \frac{2}{d(d-1)} \, A_{hv}A^*_h
\end{equation}
The simplicity of this expression (the factorization of the simplicial
spacetime and its dual) is a direct consequence of the inherent
orthogonality between the Voronoi and Delaunay lattices, and its
impact on this calculation, and in Regge calculus as a whole, cannot
be overstated. These d-cells are the Regge-calculus hybrid versions of
the reduced Brillouin cells commonly found in solid state physics,
though they are hybrid because they are coupled to their dual
structures in the underlying atomic lattice. We view these as the
fundamental building blocks of lattice gravity and at the Planck scale
perhaps the Regge calculus version of Leibniz's Monads -- {\em
  Vinculum Substantiale}. The scalar factor in this expression, which
depends on the dimension of the lattice, was derived in the appendix
of an earlier paper \cite{M97}. Furthermore we obtain the complete
proper $d$-volume, $\Delta V_v$, by linearly summing (\ref{Vhv}) over
each of the triangles $h$ in the Delaunay lattice sharing vertex $v$.
\begin{equation}
\Delta V_v = \sum_{h_{|v}} \Delta V_{hv} 
                = \frac{2}{d(d-1)} \sum_{h_{|v}} A_{hv}A^*_h
\end{equation}
We now can re-express the Regge-Hilbert action at a vertex in terms of
these hybrid blocks. 
\begin{equation}
\label{Rv}
  I_v =  \frac{1}{16 \pi} \sum_{v} R_v \sum_{h_{|v}} \frac{2}{d(d-1)} A_{hv}A^*_h 
\end{equation}

We now return to the more familiar hinge-based Regge-Hilbert action
($I_h$) of (\ref{Ih}). The proper 4-volume associated to hinge $h$ has
been shown to be factorable in terms of the area of the triangle hinge
and its corresponding dual Voronoi area \cite{M97}.
\begin{equation}
\Delta  V_h = \frac{2}{d (d-1)} A_h A^*_h.
\end{equation} 
Following the procedure discussed above, we can express the area of
$h$ a sum of its circumcentrically-partitioned pieces
(Figure~\ref{fig:ahv}).
\begin{equation}
A_h = \sum_{v_{|h}} A_{hv}
\end{equation}
Therefore the action per hinge (\ref{Ih}) can be expressed as the following
double summation:
\begin{equation}
  I_h = \frac{1}{16 \pi} \sum_h \sum_{v_{|h}} R_h \left( \frac{2}{d(d-1)} A_{hv} A^*_h\right)
\end{equation}
A key step in this derivation is the ability to switch the order of
summation, and fortunately action is unchanged if we reverse this
order.
\begin{equation}
\label{Rh}
I_h = \frac{1}{16 \pi} \sum_v \sum_{h_{|v}} R_h \left( \frac{2}{d(d-1)} A_{hv} A^*_h\right)
\end{equation} 
The vertex-based action of (\ref{Rv}) must be equal to this
hinge-based action of (\ref{Rh}). We immediately obtain the desired
expression for the Riemann scalar curvature at a vertex.
\begin{equation}
R_v = \frac{\sum_{h_{|v}} R_h A^*_h A_{hv}}
           {\sum_{h_{|v}} A^*_h A_{hv}}
= \frac{\sum_{h_{|v}} R_h A^*_h A_{hv} / \sum_{h_{|v}} A_{hv}}
       {\sum_{h_{|v}} A^*_h A_{hv}     / \sum_{h_{|v}} A_{hv}}
\end{equation}
Here we have divided the numerator and denominator by the same
quantity leaving it unchanged.  Both the numerator and denominator are
in the form of a weighted average over the "Brillion kites" ($A_{hv}$)
at vertex $v$.  We define, in a natural way, the ``kite weighted
average" at vertex $v$ of any hinge-based quantity $Q_h$ as follows:
\begin{equation}
\langle Q  \rangle_v \equiv \frac{\sum_{h_{|v}} Q_h A_{hv} }{\sum_{h_{|v}} A_{hv}}.  
\end{equation} 
Given this definition, the scalar curvature invariant at vertex $v$
can be expressed as a ``kite-weighted average'' of the integrated
hinge-based scalar curvature of Regge calculus.
\begin{equation}
R_v = \frac{\langle R_h A^*_h \rangle_v}{\langle A^*_h \rangle_v},
\end{equation}
where it was shown in \cite{M97} that the Riemann scalar curvature at
the hinge $h$ is expressible as the hinge's curvature deficit
($\epsilon_h$) per unit Voronoi area ($A^*_h$) dual to $h$.
\begin{equation}
\label{R_h}
R_h = {d(d-1)} \frac{\epsilon_h}{A^*_h}.
\end{equation}
Therefore the expression for the vertex-based scalar curvature
invariant derived here is strikingly similar to the usual Regge
calculus expression for the hinge-based scalar curvature invariant
(\ref{R_h}).  The only difference is that the numerator and
denominator of (\ref{R_h}) is replaced by their kite-weighted averages.
\begin{equation}
R_v ={d(d-1)} \frac{\langle \epsilon_h \rangle_v}
                            {\langle A^*_h      \rangle_v}
\end{equation}

\begin{figure}
\centering
\includegraphics[width=0.8\linewidth]{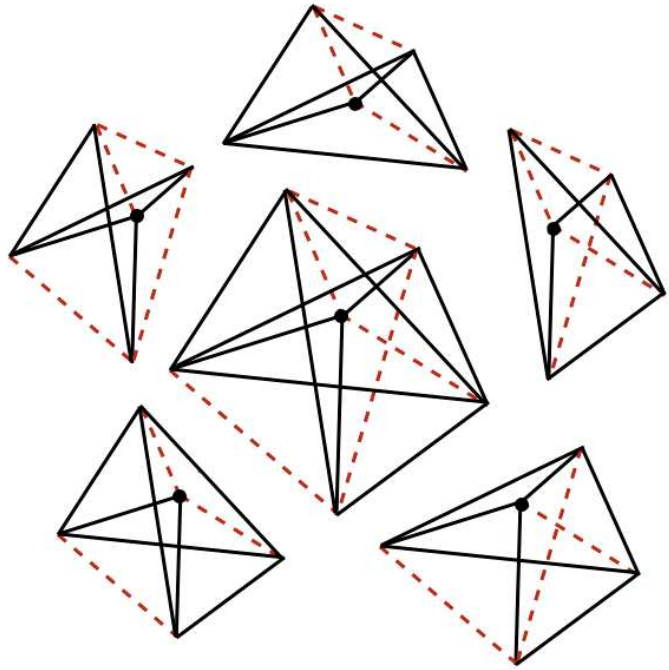}
\caption{{6-Point Scalar Curvature Detector:} The minimal 6-point
  detector at the event $v$ (depected in the picture as the black
  circle) is shown in the central region of the figure. The dashed
  lines represent null edges while the solid lines are timelike. We
  also show each of its five 4-simplices exploded off into the
  perimeter of the diagram.}
\label{fig:tlnull}
\end{figure}

In a 4-dimensional spacetime the minimum number of events needed to
measure the scalar curvature at a vertex ($v$) is six. This occurs
when the 4-dimensional Voronoi cell dual to $v$ is itself a 4-simplex.
This corresponds to the minimum allowable number of simplicies in a
Regge calculus spacetime lattice that can meet at vertex $v$ and is
consistent with earlier results on the minimum number of test
particles needed to measure the twenty components of the Riemann
tensor \cite{S60,CD86}.  Such a minimal 6-point scalar curvature
detector can be constructed solely from null (laser) and timelike
(clock) edges -- the tools available to a spacetime surveyor
\cite{MW84}.  (Figure~\ref{fig:tlnull}). Such chronometric
constructions, we believe will be useful in their applications to
discrete models of quantum gravity.

\ack We thank Renate Loll and Seth Lloyd for stimulating discussions
which provided us the motivation to continue this research, and we are
especially grateful to John A. Wheeler for providing the inspiration
and initial guidance into this field of spacetime geodesy.  We would
like to thank Florida Atlantic University's Office of Research and the
Charles E. Schmidt College of Science for the partial support of this
research.

\end{document}